\documentclass[twocolumn,pre,superscriptaddress]{revtex4}

\usepackage{dcolumn}
\usepackage{amsmath}

\usepackage{graphicx}

\newcommand{\half}{\tfrac12}
\newcommand{\etal}{{\it{}et~al.}}
\newcommand{\defn}{\textit}

\newcommand{\mat}{\mathbf}

\begin{document}

\title{Consistency of community structure in complex networks}

\author{Maria A. Riolo}
\affiliation{Santa Fe Institute, 1399 Hyde Park Road, Santa Fe, New Mexico, USA}
\affiliation{Center for the Study of Complex Systems, University of
Michigan, Ann Arbor, Michigan, USA}
\author{M. E. J. Newman}
\affiliation{Santa Fe Institute, 1399 Hyde Park Road, Santa Fe, New Mexico, USA}
\affiliation{Center for the Study of Complex Systems, University of
Michigan, Ann Arbor, Michigan, USA}
\affiliation{Department of Physics, University of Michigan, Ann Arbor,
  Michigan, USA}

\begin{abstract}
  The most widely used techniques for community detection in networks, including methods based on modularity, statistical inference, and information theoretic arguments, all work by optimizing objective functions that measure the quality of network partitions.  There is a good case to be made, however, that one should not look solely at the single optimal community structure under such an objective function, but rather at a selection of high-scoring structures.  If one does this one typically finds that the resulting structures show considerable variation, and this has been taken as evidence that these community detection methods are unreliable, since they do not appear to give consistent answers.  Here we argue that, upon closer inspection, the structures found are in fact consistent in a certain way.  Specifically, we show that they can all be assembled from a set of underlying ``building blocks,'' groups of network nodes that are usually found together in the same community.  Different community structures correspond to different arrangements of blocks, but the blocks themselves are largely invariant.  We propose an information theoretic method for discovering the building blocks in specific networks and demonstrate it with several example applications.  We conclude that traditional community detection is not the failure some have suggested it is, and that in fact it gives a significant amount of insight into network structure, although perhaps not in exactly the way previously imagined.
\end{abstract}

\maketitle

\section{Introduction}
Many networks, from social and information networks to biological networks and the internet, are found to divide into distinct groups of nodes, referred to variously as modules, clusters, or communities~\cite{GN02,Fortunato10}.  Community detection---the process of identifying such groups in unlabeled network data---is widely used as an analytical tool for exploring the large-scale structure of complex networks.  Many algorithms for community detection have been proposed, but the most widely used ones all share one feature in common: they operate by optimizing some kind of objective function that measures the quality of candidate divisions of a network into communities.  Perhaps the most widely used method is modularity maximization, as embodied for instance in the spectral modularity and Louvain algorithms, which work by optimizing the heuristic objective function known as modularity~\cite{NG04,Newman06b,BGLL08}.  Inference methods, such as methods based on the stochastic block model, work by optimizing the likelihood of the observed network under an appropriate network model~\cite{HLL83,NS01,KN11a}.  The widely used InfoMap method works by maximizing the entropy of a random walk on the network~\cite{RB07b}.

However, as pointed out by a number of authors~\cite{GSA04,MD06,CMN08}, simply reporting the single best division of a network, as defined by an objective function, misses much of the insight that is to be gained from community analysis.  In many networks, perhaps most, there are multiple divisions of the nodes that achieve high objective-function scores and in principle any of these could be the ``correct'' division of the network.  It is a crucial question whether these competing divisions are, in some sense, similar to one another or whether, conversely, they are substantially different.  If all (or most) high-scoring divisions are similar, then we may hypothesize that the community analysis is revealing some genuine underlying truth about the network: even if we don't know which of several candidate divisions is the correct one, we may still be able to draw insight from them if the candidates all tell essentially the same story.  On the other hand, if the high-scoring divisions are quite different from one another then it is harder to argue that they are meaningful.

As an example, it is known that even completely random networks, such as Erd\H{o}s--R\'enyi style random graphs, have divisions with high modularity scores~\cite{GSA04,RB06b}, yet such random networks clearly have no community structure by any reasonable definition.  Massen and Doye~\cite{MD06} generated a selection of high-modularity divisions of random graphs by Monte Carlo sampling and found that competing divisions of the same graph had little common structure, suggesting that they are probably not meaningful---a reasonable conclusion in the case of a random graph.  Subsequent theoretical work has bolstered this viewpoint using ideas borrowed from the physics of glassy systems.  If we consider the modularity as an energy function for a thermal model, then the random graph can be shown to undergo a transition with decreasing temperature to a replica symmetry broken state where there are many competing modularity maxima that correspond to essentially unrelated divisions~\cite{RB06b,RL08,DKMZ11b,ZM14}.

In many real-world networks, by contrast, as well as certain model networks such as the stochastic block model, it is believed that there is clear and meaningful community structure, which we would like to be able to extract and analyze with our algorithms.  In these cases we would hope that, to the extent that there are competing divisions with high scores, those divisions would be largely similar to one another, at least in their gross features.  Thus, the existence of true community structure would be associated with the observation that high-scoring divisions are similar and its absence with the observation that they are different.  Equivalently, true community structure would correspond to replica symmetry and lack of it to replica symmetry breaking.

Unfortunately, some previous studies have found this not to be the case.  For example, Good~\etal~\cite{GDC10} generated Monte Carlo samples of high-modularity divisions for a range of networks, including both models and real-world examples, and found in all cases that even though the networks in question were believed to possess strong community structure there were nonetheless a large number of high-scoring divisions that appeared to be quite different.  This raises serious questions about whether our community detection algorithms are returning meaningful results.

\begin{figure}
\begin{center}
\includegraphics[width=6cm]{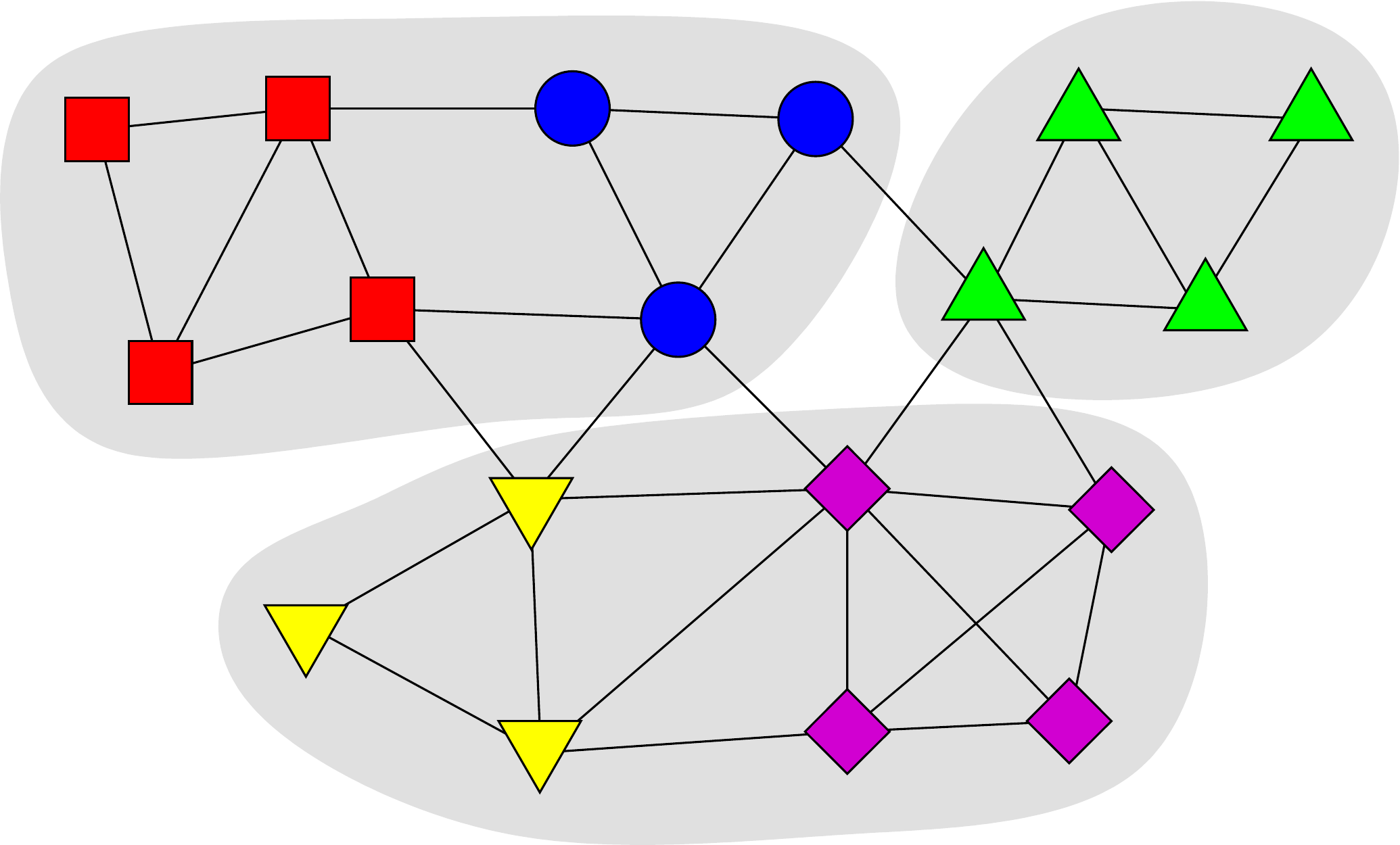} \\
\ \\
\ \\
\includegraphics[width=6cm]{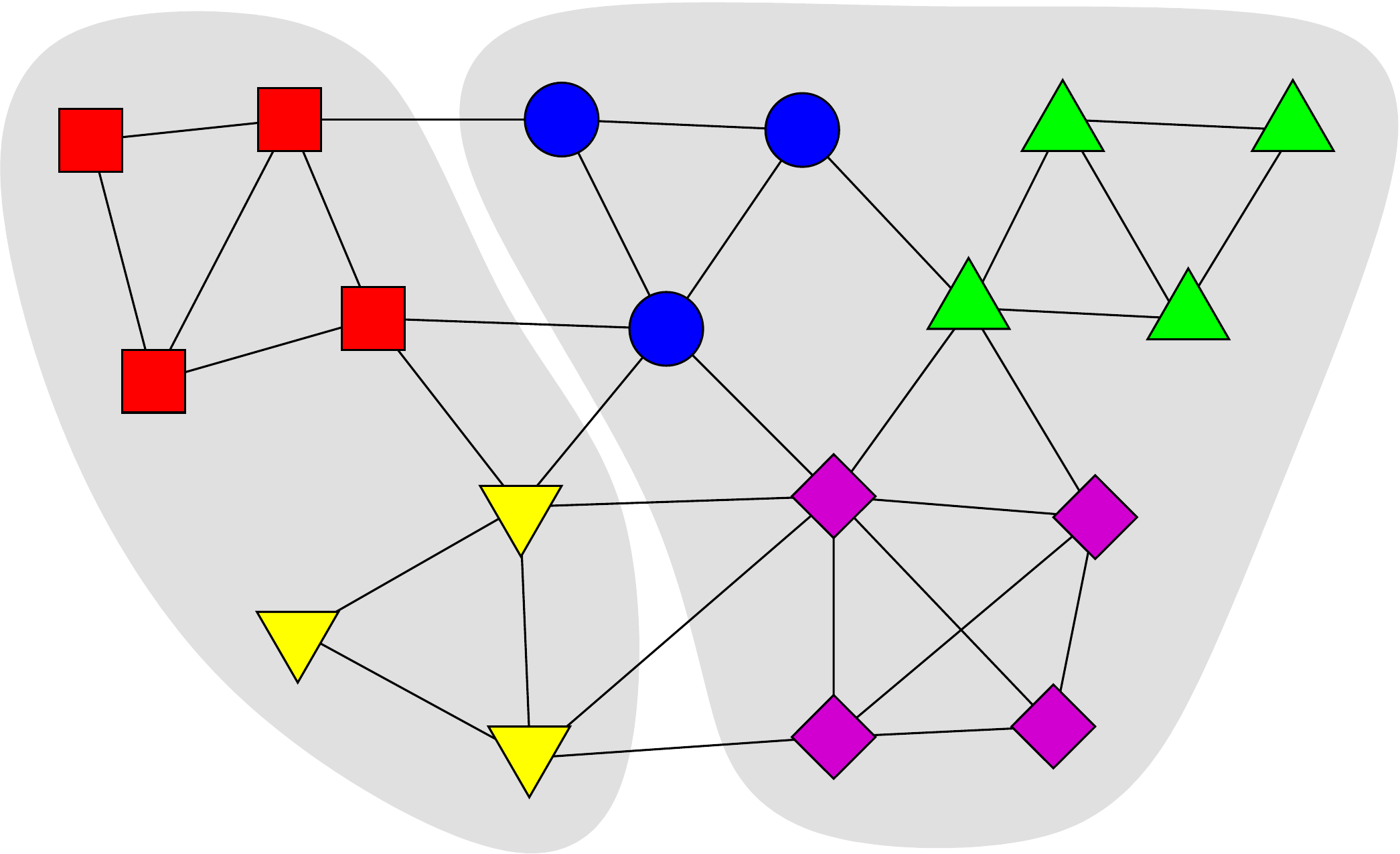}
\end{center}
\caption{Two divisions of the same set of network building blocks.  The five building blocks are denoted by the shapes and colors of the nodes and the community divisions are denoted by the shaded areas.  Each community division can be thought of as a different way of assembling the building blocks into communities.}
\label{fig:example}
\end{figure}

In this paper we revisit this question and show that in fact the high-scoring divisions of many networks \emph{are} similar, but in a more subtle sense.  Specifically, we show that while it is true that the communities discovered by these algorithms vary substantially between high-scoring divisions, the variation is of a limited and specific type.  We show that for both real and model networks it is possible to find an elemental set of ``building blocks,'' groups of nodes such that most high-scoring community divisions are formed by combining these blocks in one way or another, while the blocks themselves are essentially indivisible---see Fig.~\ref{fig:example} for a sketch.  Thus most high-scoring community divisions are similar in the sense of being built from the same set of building blocks.

To put this another way, if we know the blocks then it takes very little additional information to specify how they are joined together and hence specify the complete community structure.  We use this observation to create an information-theoretic algorithm for determining the building blocks and demonstrate its use on a range of example networks.

Our conclusion from these findings is that community structure analyses do in fact convey consistent and believable information about the large-scale structure of networks, when interpreted in an appropriate manner.

\section{Sampling network divisions}
\label{sec:mc}
Like the previous studies discussed above, our investigation starts with the generation of a random sample of network divisions that score highly according to an appropriate objective function.  Previous studies sampled divisions according to modularity, but this approach is arguably somewhat ad hoc: there is no rigorous principle that tells us the relative sampling weight one should give to divisions with different modularity.    Massen and Doye and others~\cite{MD06,RB06a,GDC10,ZM14} have employed a Boltzmann distribution, which is convenient for numerical simulation but does not have a formal justification in this context.  In our work we use an alternative approach that has become popular in recent years, that of sampling from the posterior distribution of an appropriate generative model.  The model we use, which is standard in calculations of this kind, is the degree-corrected stochastic block model~\cite{KN11a}, a~random graph model in which the probabilities of edges depend on the communities they belong to.  Inverting the probability relation using Bayes' rule allows us to write an expression for the probability of a particular community division given an observed network and it is from this distribution that we sample.  Specifically, the approach is as follows.  (This part of the paper follows the outline of our previous presentation in~\cite{RCRN17}---see that paper for additional details.)

The degree-corrected stochastic block is a random generative model of a community-structured network.  When used to generate networks (rather than for community detection), it works as follows.  Initially, each of $n$ nodes is assigned to one of $k$ groups, then a Poisson-distributed number of edges is added between each node pair such that the expected number of edges between nodes~$i$ and~$j$ is $\theta_i \theta_j \omega_{g_ig_j}$, or a half this many when $i=j$, where $\theta_i$ and $\omega_{rs}$ are parameters that we choose and $g_i$ is the community to which node~$i$ belongs.  This leaves $\theta_i$ and $\omega_{rs}$ arbitrary to within multiplicative constants, which are fixed by normalizing the~$\theta_i$ such that their mean is~1 in each community thus:
\begin{equation}
{1\over n_r} \sum_{i=1}^n \theta_i \delta_{r,g_i} = 1,
\label{eq:normalization}
\end{equation}
where $\delta_{ij}$ is the Kronecker delta and $n_r = \sum_i \delta_{r,g_i}$ is the number of nodes in group~$r$.

Now consider a specific undirected network with structure described by its adjacency matrix~$\mat{A}$ having elements~$A_{ij}=1$ if there is an edge between nodes~$i$ and~$j$ and 0 otherwise.  The probability, or likelihood, that this network is generated by the degree-corrected stochastic block model is
\begin{align}
P(\mat{A}|\theta,\omega,g,k) &= \prod_{i<j} \bigl( \theta_i\theta_j\omega_{g_ig_j} \bigr)^{A_{ij}} e^{-\theta_i\theta_j\omega_{g_ig_j}} 
    \nonumber\\
   &\qquad{}\times
     \prod_i \bigl( \half\theta_i^2\omega_{g_ig_i}
     \bigr)^{A_{ii}/2} e^{-\theta_i^2\omega_{g_ig_i}/2} \nonumber\\
  &\hspace{-6em}{} = \prod_i \theta_i^{d_i} 
  \prod_{r<s} \omega_{rs}^{m_{rs}} e^{-n_r n_s\omega_{rs}}
  \prod_r \omega_{rr}^{m_{rr}} e^{-n_r^2\omega_{rr}/2},
\label{eq:likelihood}
\end{align}
where we have used Eq.~\eqref{eq:normalization} in the second equality, $d_i = \sum_j A_{ij}$ is the degree of node~$i$, and
\begin{equation}
m_{rs} = \left\lbrace\begin{array}{ll}
         \sum_{ij} A_{ij} \delta_{g_i,r} \delta_{g_j,s} &
         \qquad\mbox{when $r\ne s$,} \\
         \rule{0pt}{15pt}\half \sum_{ij} A_{ij} \delta_{g_i,r} \delta_{g_j,r} &
         \qquad\mbox{when $r=s$,}
         \end{array}\right.
\end{equation}
which is the number of edges between groups $r$ and~$s$.  We have also neglected an overall multiplying factor in Eq.~\eqref{eq:likelihood} which is independent of the parameters $\theta$, $\omega$, and~$g$, and will therefore have no effect on our calculations.

The values of the parameters~$\theta$ and~$\omega$ are not of interest in the present case, so we integrate them out using maximum-entropy priors as described in~\cite{RCRN17}, to get
\begin{align}
P(\mat{A}|g,k) &= \prod_r n_r^{\kappa_r} {(n_r-1)!\over(n_r+\kappa_r-1)!}
   \nonumber\\
  &{}\times \prod_{r<s} {m_{rs}!\over(pn_r n_s+1)^{m_{rs}+1}}
     \prod_r {m_{rr}!\over (\half p n_r^2+1)^{m_{rr}+1}},
\label{eq:pagk}
\end{align}
where
\begin{equation}
\kappa_r = \sum_i d_i \delta_{r,g_i}
\end{equation}
and we have discarded a further multiplying constant.  Now we apply Bayes' rule to get
\begin{equation}
P(g,k|\mat{A}) = {P(\mat{A}|g,k) P(g,k)\over P(\mat{A})}.
\end{equation}
The denominator $P(\mat{A})$ is a simple normalizing constant that plays the role of a partition function and, like other constants, will not be important for our calculations.  For the prior probability~$P(g,k)$ we again follow our previous work, making the choice $P(g,k) = n^{-k} \prod_r n_r!$, which is derived from a simple ``restaurant process''~\cite{RCRN17}.  With this choice, and again neglecting overall constants, we have
\begin{align}
P(g,k|\mat{A}) &= n^{-k}
   \prod_r n_r^{\kappa_r} {n_r! (n_r-1)!\over(n_r+\kappa_r-1)!}
   \nonumber\\
  &{}\times \prod_{r<s} {m_{rs}!\over(pn_r n_s+1)^{m_{rs}+1}}
     \prod_r {m_{rr}!\over (\half p n_r^2+1)^{m_{rr}+1}}.
\label{eq:pgka}
\end{align}

We now generate community divisions~$(g,k)$ from this distribution using Metropolis--Hastings Monte Carlo sampling.  Our sampling algorithm, which makes specific use of the structure of the prior on $g$ and $k$ to enhance sampling speed, is described in detail in~\cite{RCRN17}.  The implementation, which is written in the C programming language, performs about 1 million Monte Carlo steps per second on a typical desktop computer, allowing our calculations to scale to networks of tens or hundreds of thousands of nodes with relative ease, although we will have no need of such large networks in this paper.

\section{Results}
Our goal is to use the algorithm described above to generate a random sample of high-probability community divisions and then compare the structure of those divisions to try to determine what features they have in common.  As described in the introduction, we find that in most cases they can be represented as the union of a collection of elemental and largely indivisible blocks of nodes that appear to represent the fundamental ``atoms'' of community structure in the network.

\subsection{An example model network}
\label{sec:cliques}
\begin{figure}
\begin{center}
\includegraphics[width=7cm]{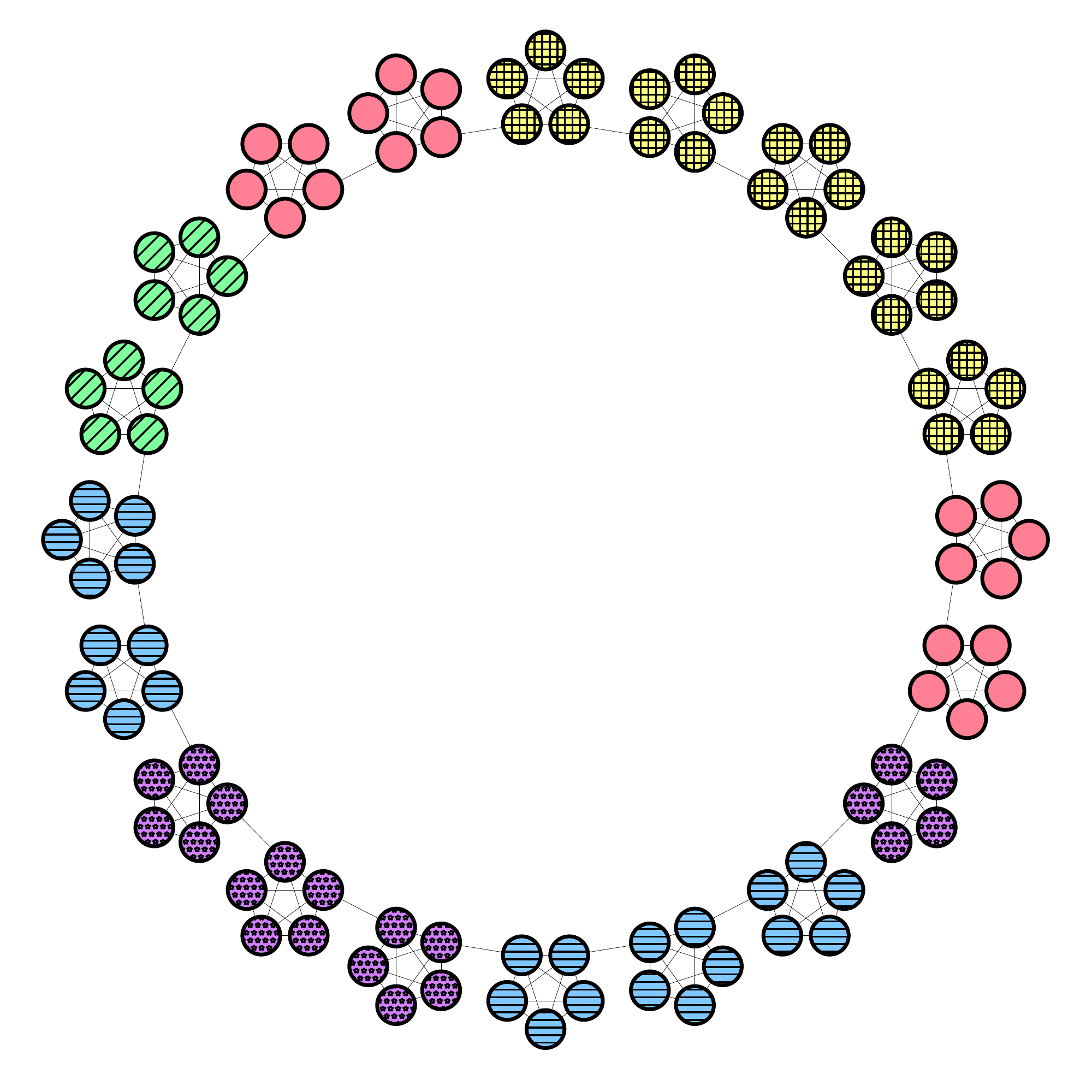}
\end{center}
\caption{The model network network of Good~\etal~\cite{GDC10} with nodes colored to indicate the highest probability division found over 20 runs of $10^7$ steps of the Monte Carlo algorithm described in Section~\ref{sec:mc}.}
\label{fig:ring}
\end{figure}

To illustrate this approach we take for our first example a simple model network proposed by Good~\etal~\cite{GDC10} precisely as an illustration of the shortcomings of conventional community detection.  This network, which is illustrated in Fig.~\ref{fig:ring} and is similar to the ``connected caveman'' model of Watts~\cite{Watts99a}, is composed of a number of cliques (i.e.,~completely connected subgraphs) joined together in a ring.  In the example shown in the figure there are 20 cliques of five nodes each.

We now apply our Monte Carlo sampling algorithm to this network.  Figure~\ref{fig:ring} shows the highest probability division found over twenty runs of the algorithm.  The division is perhaps not the one we would at first guess---it is not the division into the 20 cliques themselves.  Instead, as the figure shows, the algorithm has divided the network into five groups of varying size.  The cliques themselves are still intact---none of them has been split between communities---but some cliques have been joined together to make larger communities of 10, 20, or even 25 nodes.

\begin{figure*}
\begin{center}
\includegraphics[width=8cm]{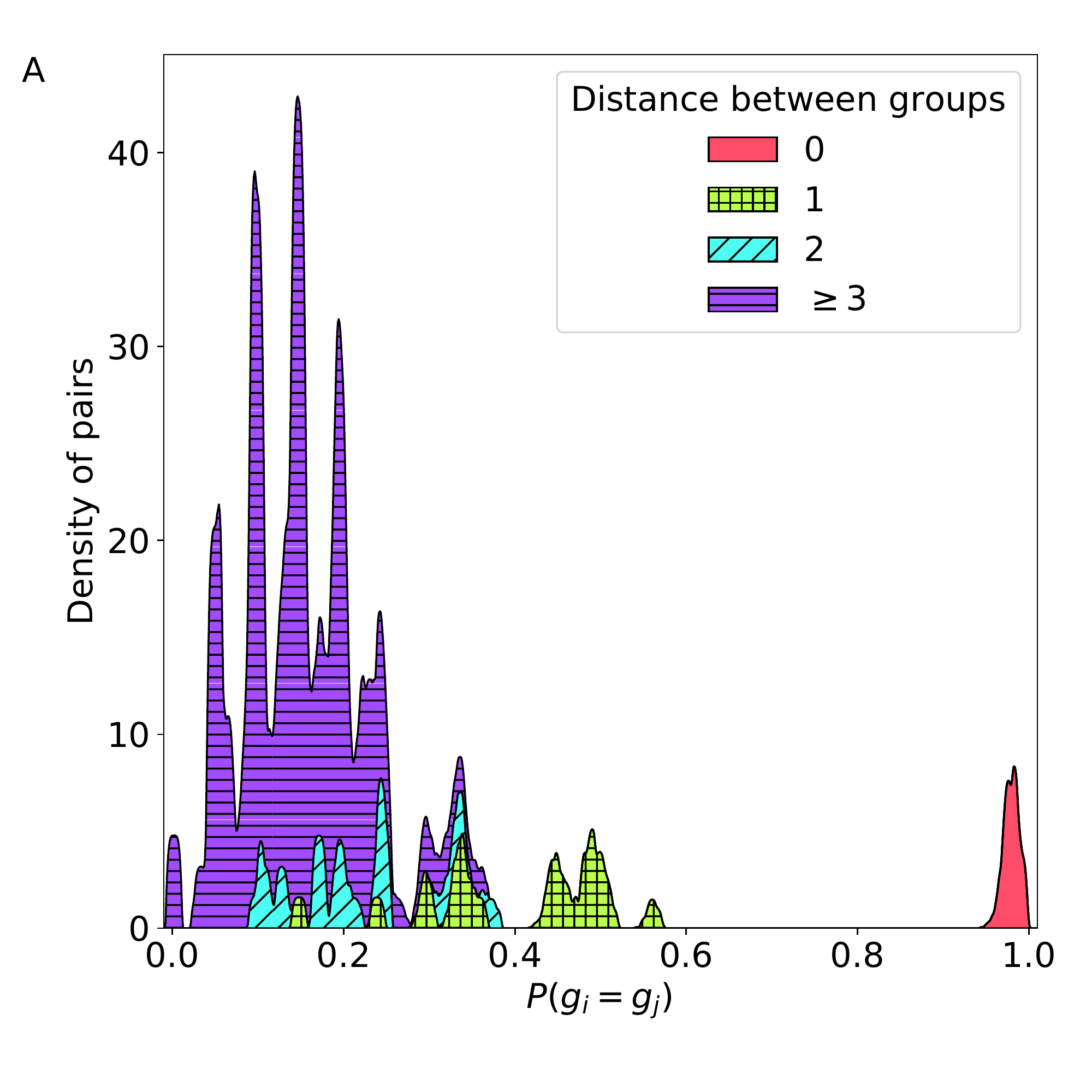}\hfill
\includegraphics[width=8.8cm]{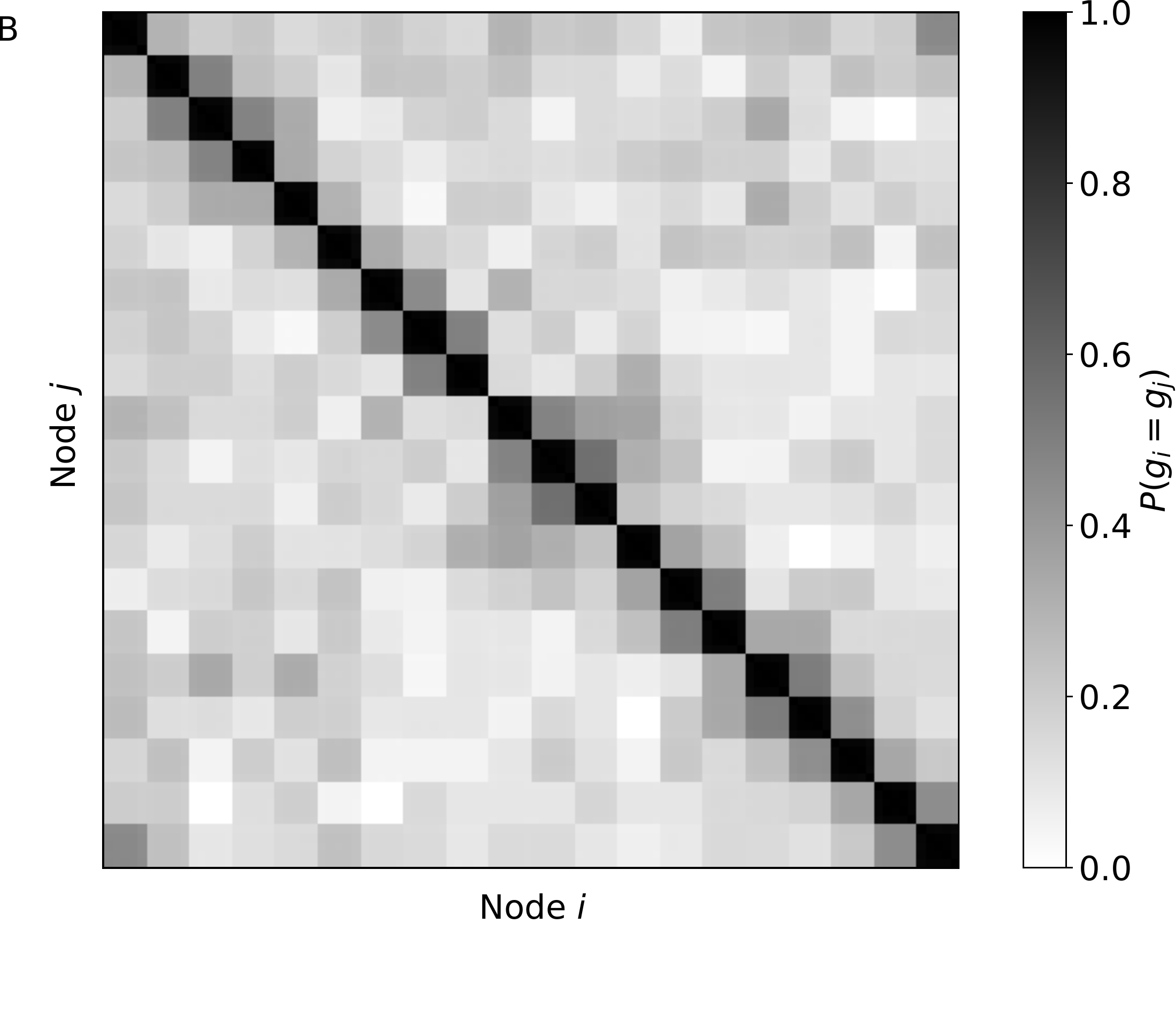}
\end{center}
\caption{(A)~The distribution of the probabilities $P(g_i=g_j)$ that two nodes~$i$ and~$j$ are found to be in the same community.  We show separate histograms for node pairs in the same clique (``distance~0''), adjacency cliques (``distance~1''), and so forth.  (B)~A density plot of the same set of probabilities.  The ground-truth cliques in the network are clearly visible as the dark squares along the diagonal.}
\label{fig:together}
\end{figure*}

As we will see, this result is typical.  There are natural blocks of nodes in many networks that want to be in the same community---the cliques in this case---but these blocks are, in most cases, not themselves communities.  The communities are assembled by putting blocks together.  Moreover, it is easy to see in this case that there are many ways of putting the blocks together that are as good as the one shown in Fig.~\ref{fig:ring}, or nearly so.  For instance, since the network has a discrete rotational symmetry around the ring, there are trivially 20 rotational variants of the division shown that have the exact same probability but which join the blocks in different ways.  The result is that if one samples many high-scoring divisions of the network one will see the same blocks repeatedly but not necessarily the same communities.  Indeed the communities can change dramatically from one state of the sampling algorithm to another: large pieces can shear off and form their own community, or join another.  If one were to compare different community divisions, therefore, particularly using elementary numerical measures of similarity such as the Rand index, one might conclude that there was wide variation between divisions and little consistency---and hence that the algorithm was not giving useful information about network structure.  This, however, would be a mistake.  Once we understand the nature of the building blocks from which the communities are assembled we see that the structures sampled by the algorithm are in a sense highly similar and consistent.

One way to make these observations more quantitative is illustrated in Fig.~\ref{fig:together}.  In panel~A of the figure we demonstrate that the individual cliques in the network are rarely split between communities.  The plot shows a histogram of the probability that each pair of nodes in the network find themselves in the same community, averaged over a large number of divisions of the network sampled using the Monte Carlo algorithm.  The histogram is colored according to the distance between cliques, where distance~0 means node pairs in the same clique, distance~1 means adjacent cliques, and so forth.  As we can see, nodes in the same clique have probability close to 1 of being in the same community, but nodes at all other distances have substantially lower probability.

In Fig.~\ref{fig:together}B we show another representation of the same probability measurements, a density plot of the pairwise probabilities.  This plot clearly picks out the individual building blocks of the network as the dark squares along the diagonal of the figure.  If we did not already know what the blocks were for this network, we could deduce them by examining this figure.

\subsection{Social network}
\begin{figure*}
\begin{center}
\hfill\includegraphics[width=15cm]{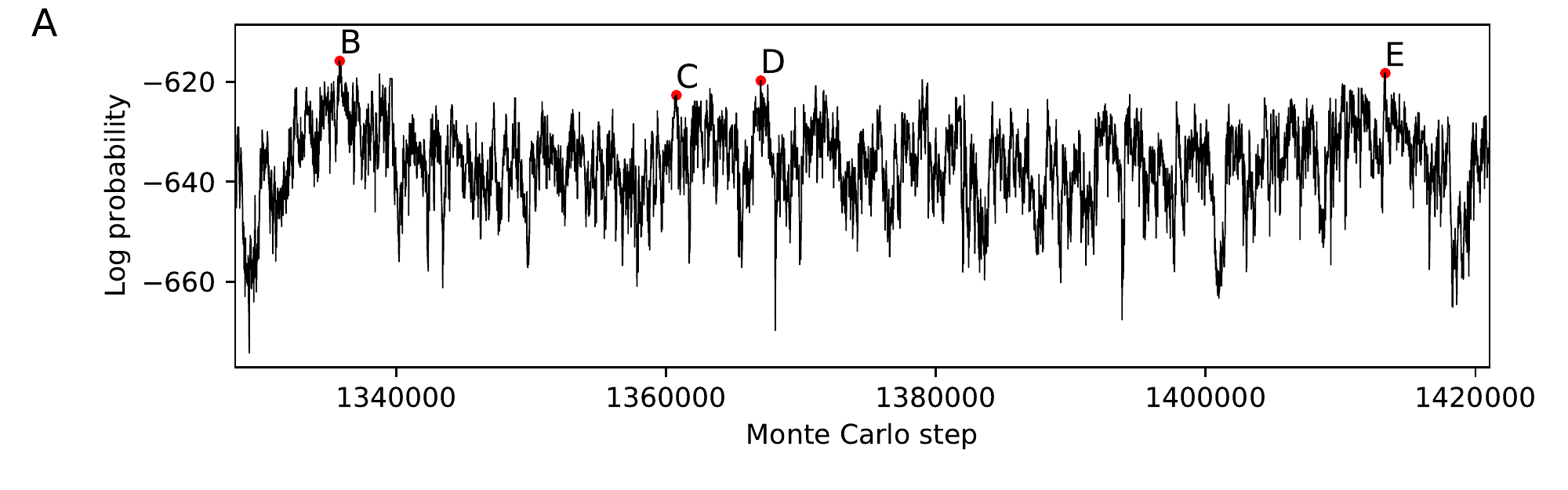}\hfill\null\\
\ \\
\ \\
\hfill\includegraphics[width=6.5cm]{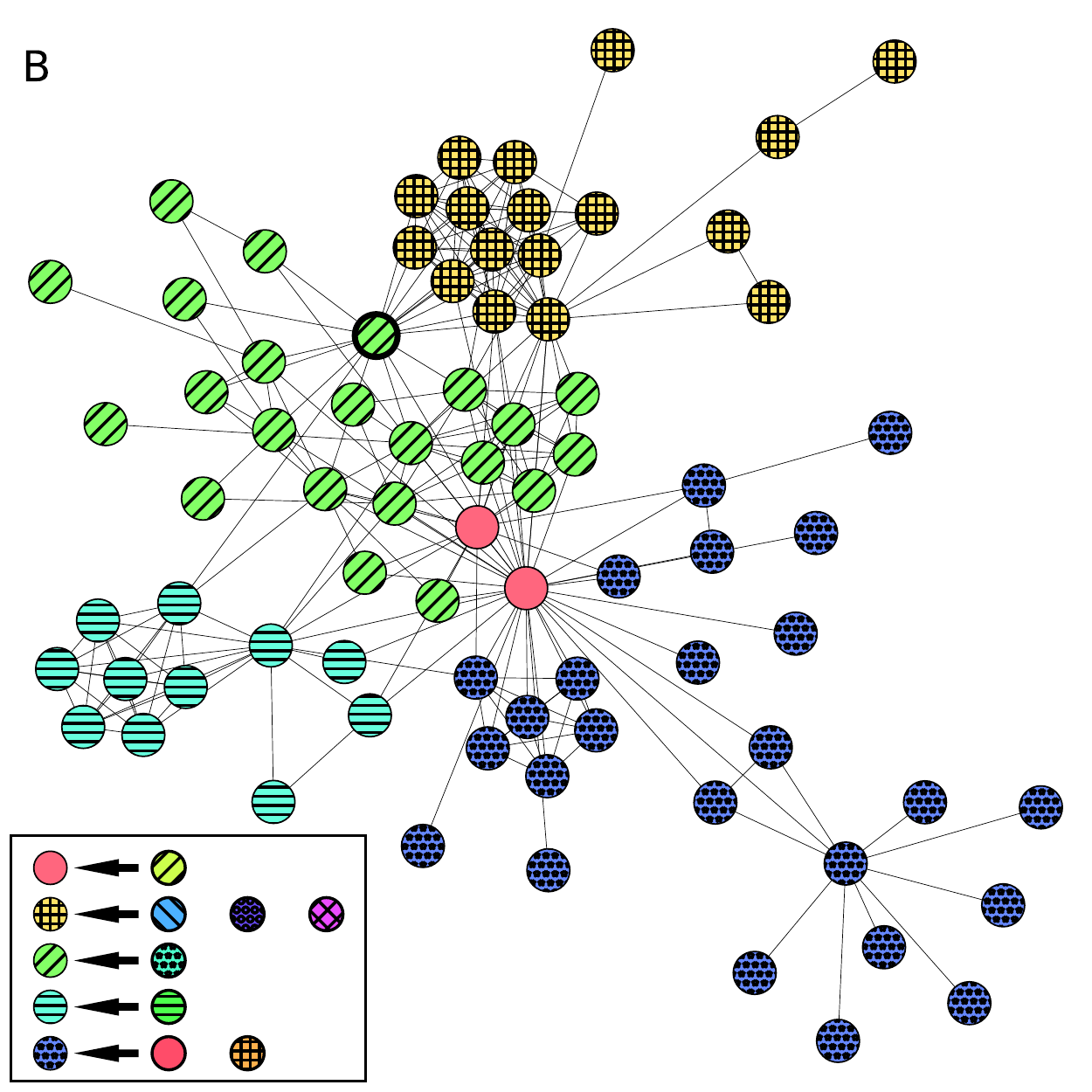}\hfill
\includegraphics[width=6.5cm]{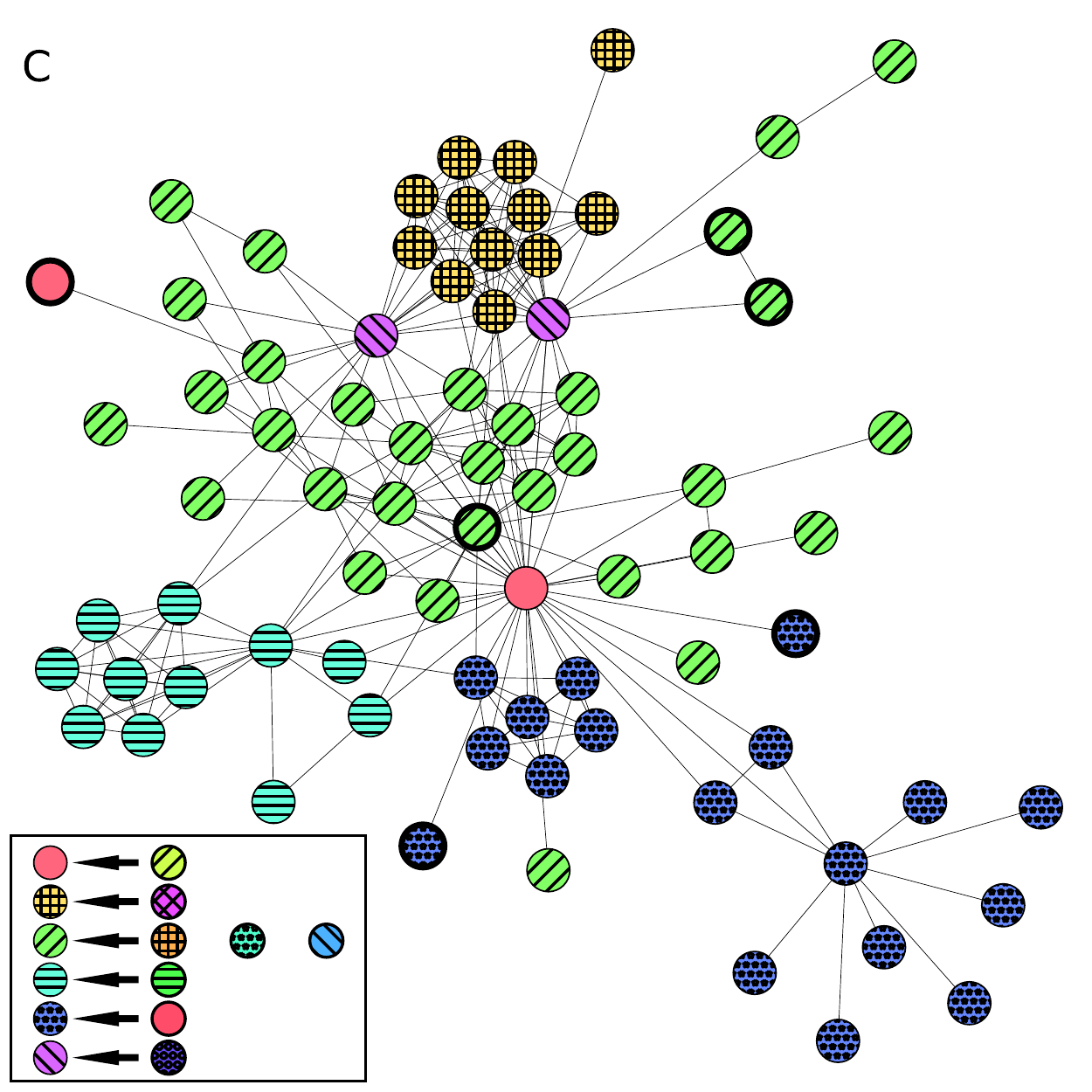}\hfill\null\\
\ \\
\ \\
\hfill\includegraphics[width=6.5cm]{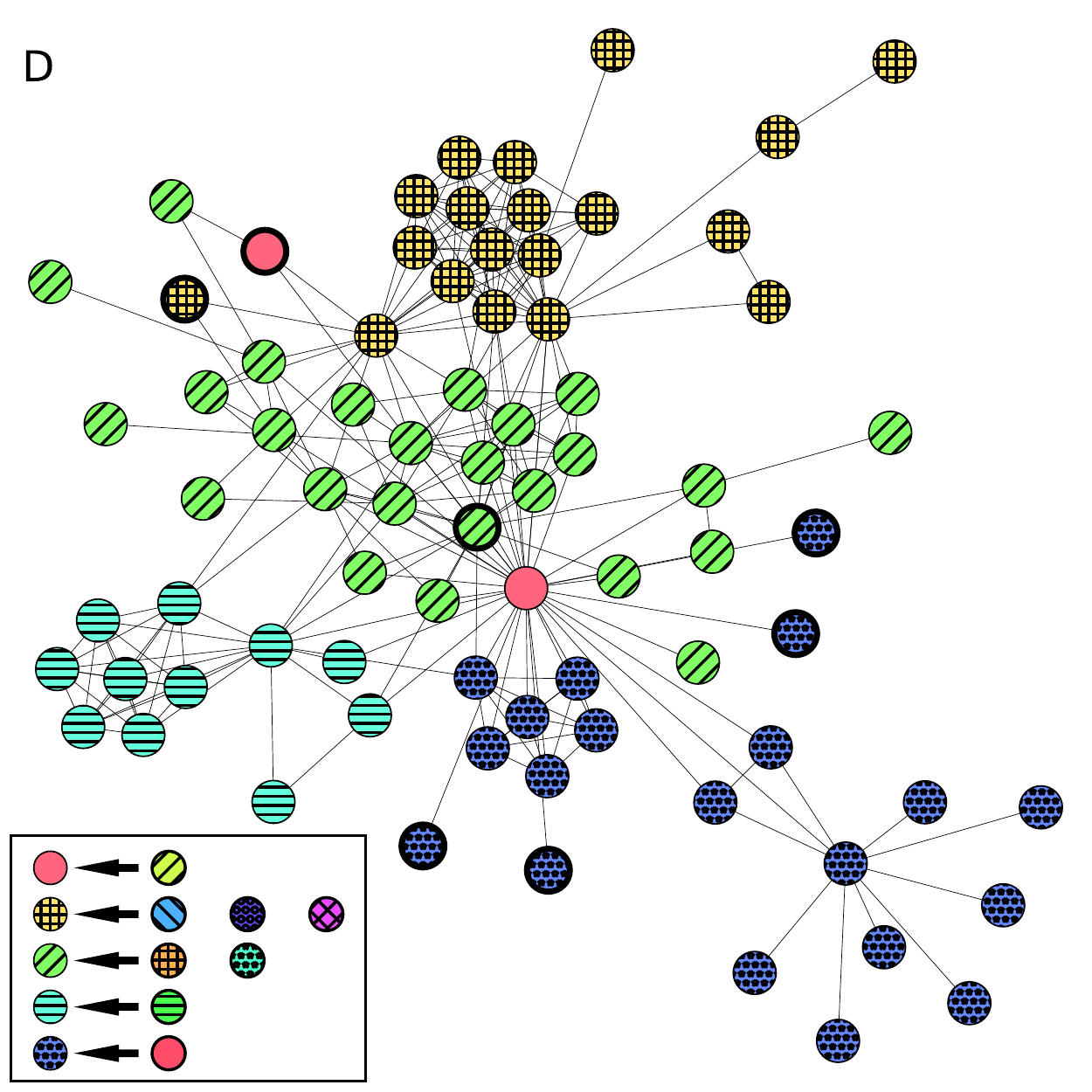}\hfill
\includegraphics[width=6.5cm]{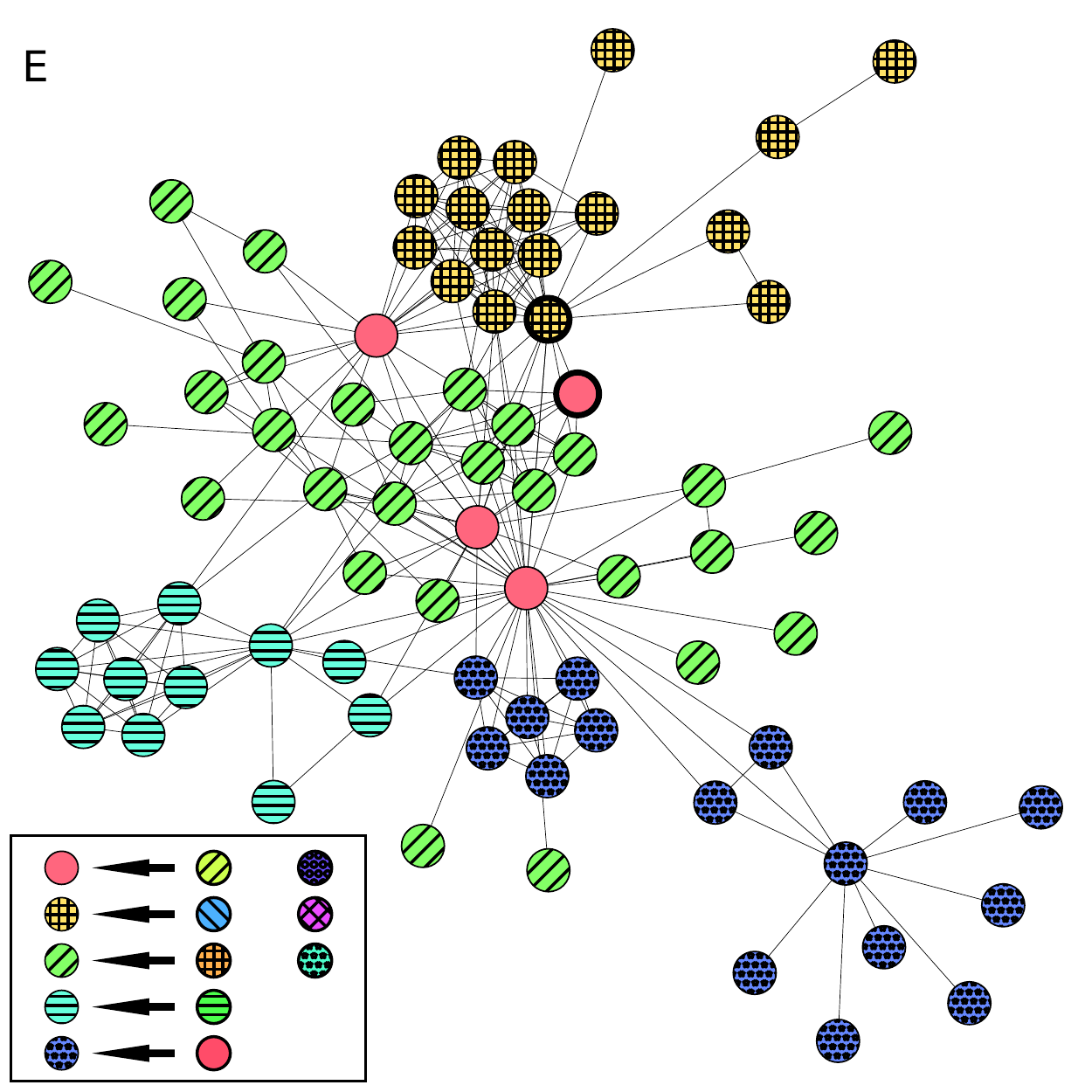}\hfill\null\\
\end{center}
\caption{Results from a single run of the Monte Carlo community detection algorithm on the fictional social network from \textit{Les Mis\'erables} (\cite{Knuth93}). In panel A, we show the log likelihood of states visited as a function of time for a portion of the run.  Selected peaks in likelihood are labeled B to~E and the community assignments at these peaks are shown in the lower four panels.  Inset in each of these four panels is a legend showing how the communities discovered by the algorithm correspond to the building blocks shown in Fig.~\ref{fig:MI}.  In each panel a small number of nodes are highlighted.  These are incorrectly assigned by the building block decomposition.}
\label{fig:lesmis}
\end{figure*}

Let us now apply the same ideas to a more complex example.  In Fig.~\ref{fig:lesmis} we show the results of applying the algorithm of Section~\ref{sec:mc} to a standard and widely studied network from the community detection literature, the social network of fictional characters in the novel \textit{Les Mis\'erables} by Victor Hugo~\cite{Knuth93}, which provides a good illustration of the phenomena discussed above.  As an initial test of the method we perform a single run of our algorithm for over a million Monte Carlo steps and then select four high-probability states from the latter portion of the run as shown in Fig.~\ref{fig:lesmis}A.  The first state has the highest probability of the four, but the others are also competitive.  Panels~B to E in Fig.~\ref{fig:lesmis} show the community divisions found in each of the four cases.

\begin{figure*}
\begin{center}
\hfill\includegraphics[width=6.5cm]{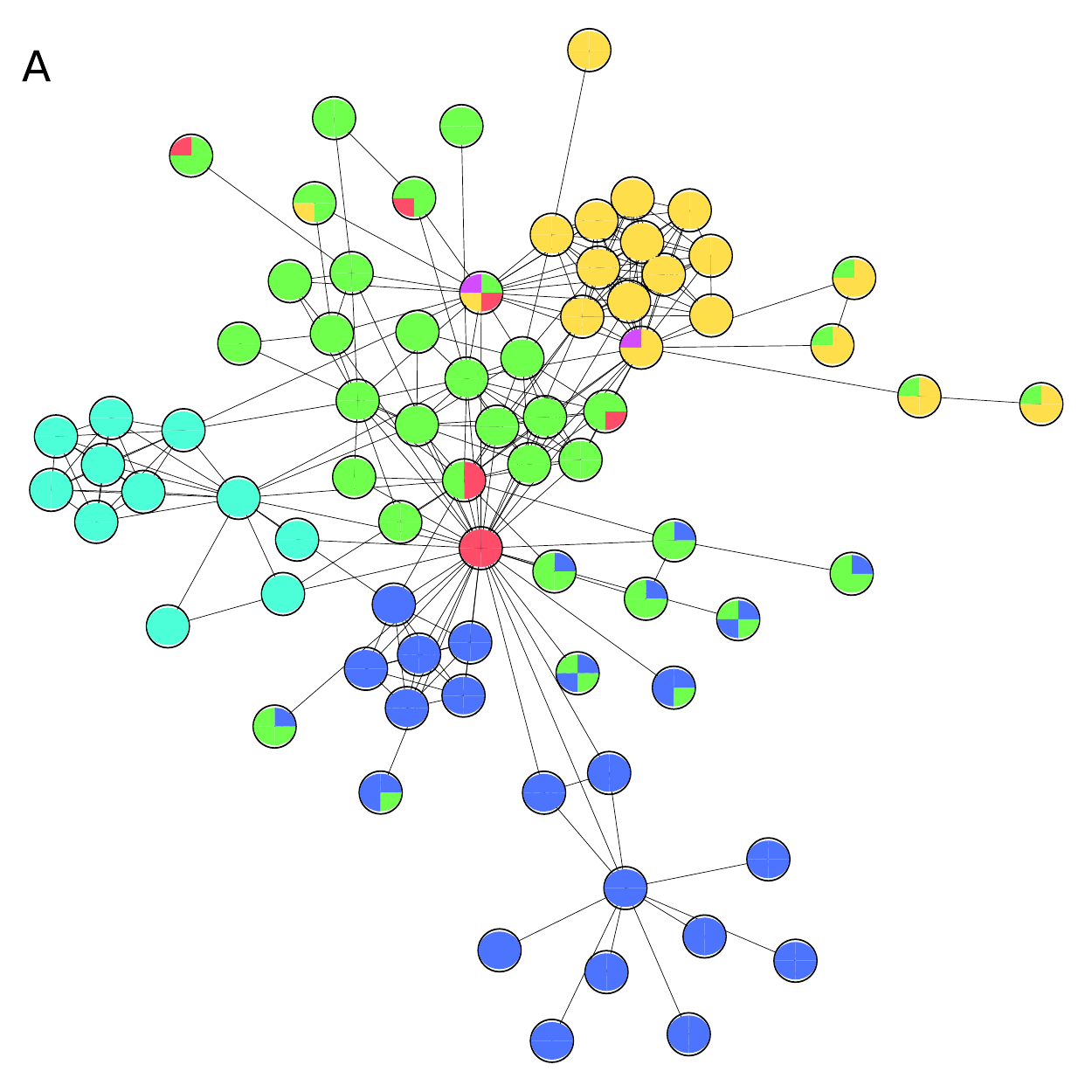}\hfill
\includegraphics[width=6.5cm]{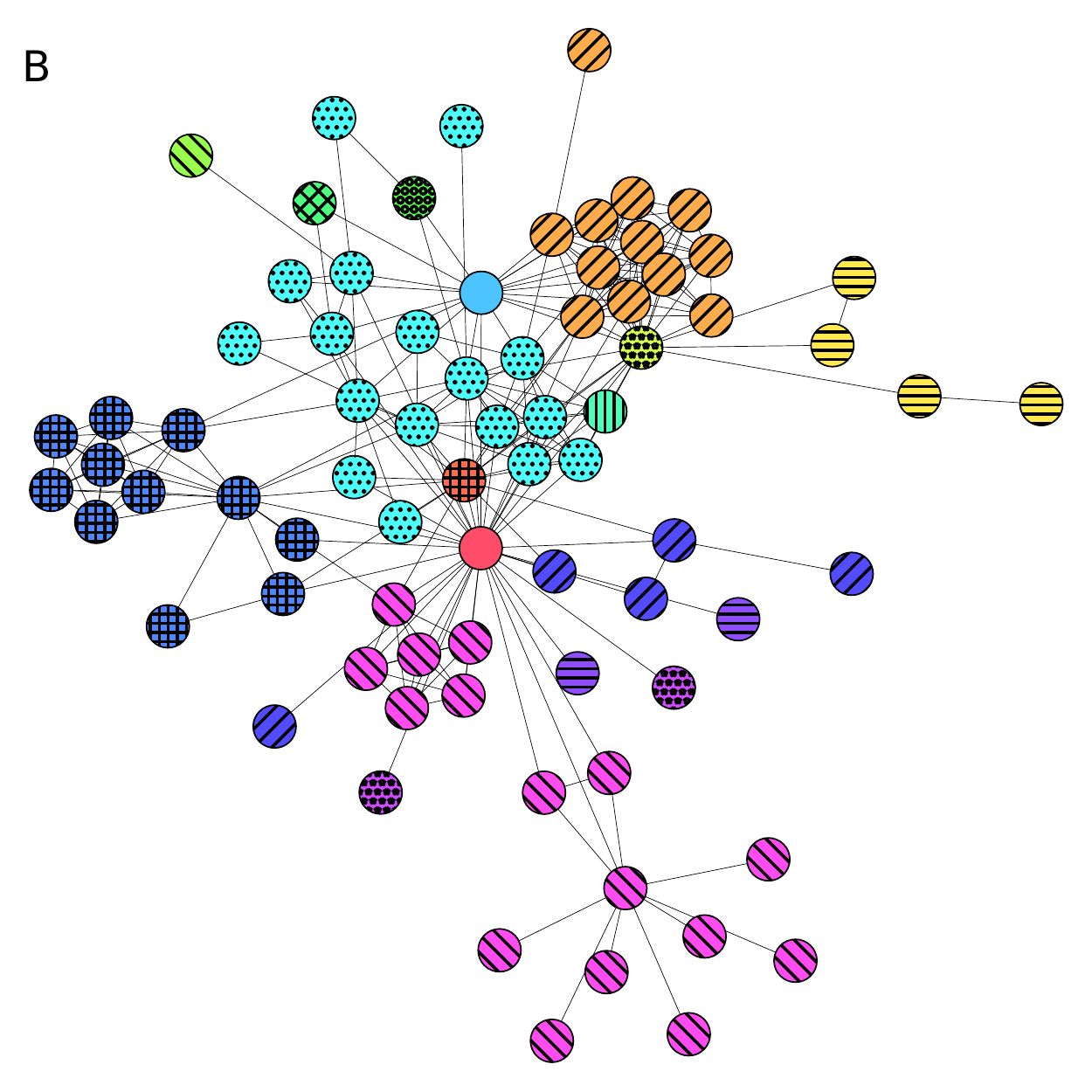}\hfill\null
\end{center}
\caption{Comparison of the four divisions of the \textit{Les Mis\'erables} social network pictured in Fig.~\ref{fig:lesmis}.  In (A) the nodes are colored in pie-chart fashion to indicate which community they belong to in each division; In (B) we assign new group labels so that only nodes that belonged to the same group in all four assignments are grouped together.  This gives us a simple estimate of the identity of the building blocks that make up the community divisions.}
\label{fig:intersect}
\end{figure*}

As we can see from the figure, the four divisions have much in common, but there are also substantial changes from one to another.  Groups of nodes break off communities from one panel to the next and join others, in a manner similar to that of the previous section.  To get a clearer picture of these changes we show in Fig.~\ref{fig:intersect} two different comparisons of the four divisions.  In Fig.~\ref{fig:intersect}A we show the network with nodes colored in pie-chart fashion to indicate which communities they belong to in each of the four divisions.  In Fig.~\ref{fig:intersect}B we perform a simple reconstruction of the building blocks of the network by assigning a different color to each group of nodes that are always found together in the same group.  Each of our four divisions is composed of combinations of these elemental groups, and yet these groups are not themselves communities.  Consider, for example, the group of nodes in Fig.~\ref{fig:intersect}B that are colored yellow with horizontal stripes.  This group does not form a stand-alone community in any of the four divisions pictured in Fig.~\ref{fig:lesmis}B--E, yet whatever group they fall in they are always found together.

This approach for reconstructing the building blocks is, however, somewhat ad hoc.  Ideally we would prefer a more rigorous method.  We describe such a method in the next section.

\subsection{Choosing optimal building blocks}
\label{sec:MI}
How can we make the notion of building blocks for community structure more rigorous?  One approach is to think in terms of information content.  A good set of building blocks is one that describes most of the structure in most commonly occurring community divisions, meaning that given the building blocks only a small amount of additional information is needed to define a division.  For instance, we could describe a community division by first specifying to which community each block belongs and then specifying a (hopefully small) set of corrections to the resulting division for any nodes that we put in the wrong community.

Information theory tells us that in general the amount of information needed to specify the community structure given the blocks is equal to the conditional entropy of the latter given the former and we may consider a particular set of building blocks to be ``good'' if the conditional entropy is small when averaged over the distribution of community divisions, Eq.~\eqref{eq:pgka}, or (more practically) a suitable set of divisions sampled using Monte Carlo.  As we will see, it is indeed possible to find such building blocks.

In practice it is conventional to use not the conditional entropy for comparing network divisions but the mutual information, which is a simple linear transform of the conditional entropy that inverts the information scale so that divisions are maximally similar for maximum mutual information.  In our calculations we make use of the ``reduced mutual information'' of Ref.~\cite{NCY19}, which includes a correction term that allows for accurate computation of the amount of information even in cases where, as here, the number of blocks may be very different from the number of communities.  The details are as follows.

Consider a network of $n$ nodes and a specific community division of that network with $k$ communities, and suppose that we have $q$ building blocks.  Let $g_i$ represent the community to which node~$i$ belongs as previously, and let $h_i$ represent the building block.   We define a \defn{contingency table}, which is the matrix of elements~$c_{gh}$ equal to the number of nodes that belong to community~$g$ and block~$h$.  Then the \defn{reduced mutual information} is given~by
\begin{equation}
M = {1\over n} \log {n! \prod_{gh} c_{gh}!\over \prod_g a_g! \prod_h
     b_h!} - {1\over n} \log \Omega,
\label{eq:mi}
\end{equation}
where $a_g = \sum_h c_{gh}$ and $b_h = \sum_g c_{gh}$ are the row and column sums of the contingency table and $\Omega$ is the number of distinct possible contingency tables that have these row and column sums.  The term in $\log \Omega$ represents the amount of information needed to determine which community each block belongs to, and prevents us from achieving high mutual information scores simply by introducing a large number of blocks.  The optimal choice of building blocks is the one that maximizes~\eqref{eq:mi} when averaged over community structures, in other words the one that contains maximal information about those structures on average.

In the limit where we have only a single building block, we have $b_h = n$ and $c_{gh} = a_g$, so the first term in~\eqref{eq:mi} is equal to $\log1$, and vanishes.  At the same time there is only one possible contingency table, so $\Omega=1$ and the second term vanishes also.  Thus we have $M=0$ in this limit.  At the other extreme, where every node is its own building block, we have $c_{gh}=0$ or 1 for all $g,h$, and $b_h=1$.  Hence the first term in~\eqref{eq:mi} is equal to $\log (n!/\prod_g a_g!)$.  At the same time, the number of contingency tables is equal the number of ways one can assign the $n$ one-node blocks to communities of the given sizes~$a_g$, which is given by the multinomial coefficient $\Omega = n!/\prod_g a_g!$.  Hence in this case the two terms in~\eqref{eq:mi} are exactly equal to one another and again we have $M=0$.  For all other choices of blocks we expect $M$ to take a value larger than zero, so somewhere in between the two limits lies the optimal choice of building blocks.

Our goal is to maximize the Monte Carlo average of Eq.~\eqref{eq:mi} over possible choices of the blocks.  Exhaustive maximization is impractical in most cases because the number of choices is exponentially large in the size of the network, so instead we use an approximate greedy algorithm, which in practice seems to work well.  The algorithm starts with every node in a block on its own, giving $M=0$, then joins together the two blocks that most increase (or least decrease) the value of~$M$.  We repeat this process, joining blocks in pairs until all blocks have been joined into one and the value of $M$ is once again zero.  The intermediate state that we pass through with the largest value of $M$ is then taken to be our block division for the network.

\begin{figure}
\begin{center}
\includegraphics[width=\columnwidth]{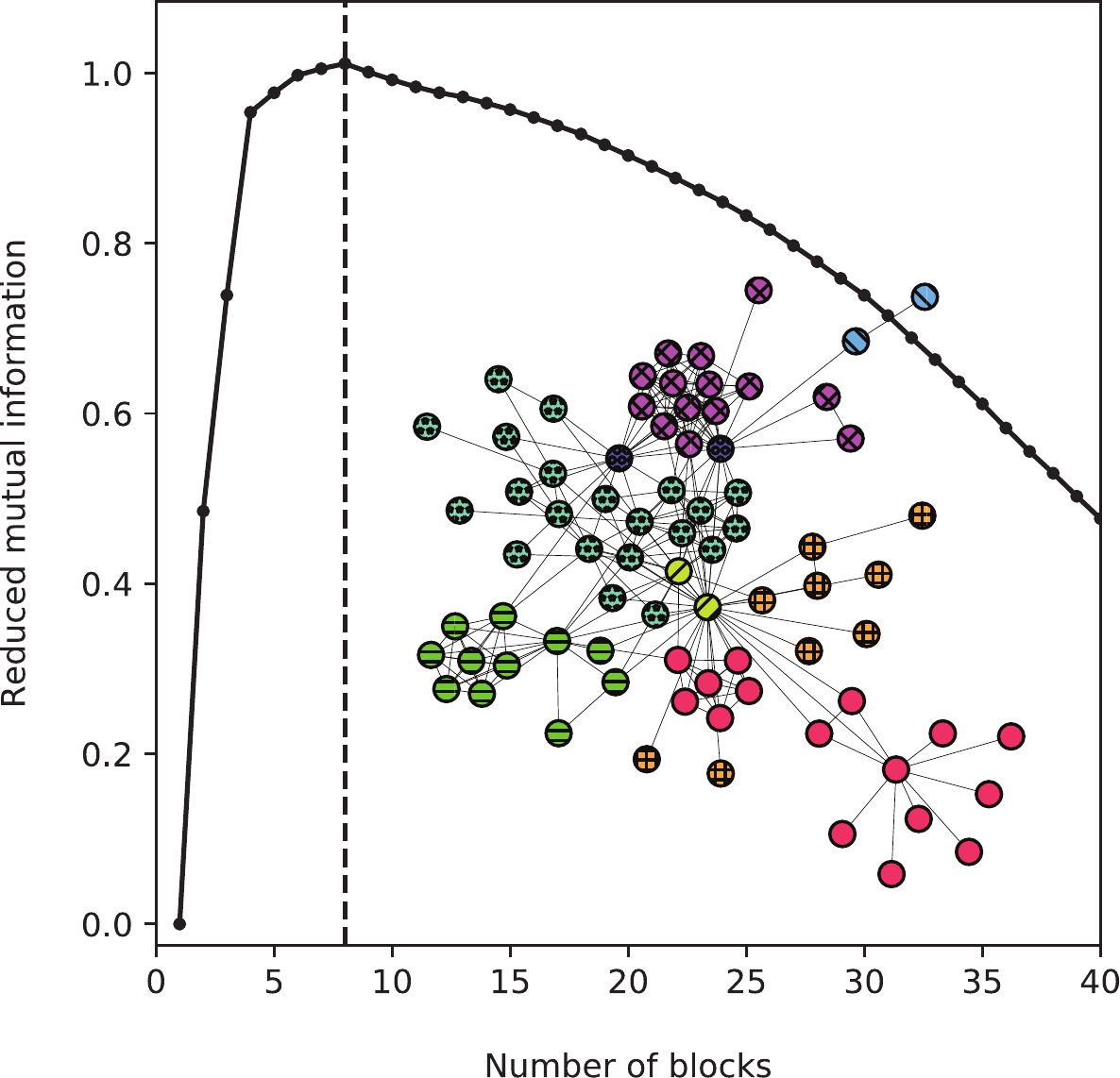}
\end{center}
\caption{Main figure: The reduced mutual information, which describes the average information about communities that is contained in the building blocks, for various different numbers of blocks, averaged over $10\,000$ sampled community structures for the \textit{Les Mis\'erables} network.  As expected, the value is small for the case of very many blocks, or very few, but there is an intermediate maximum, in this case at eight blocks, where the blocks contain the most complete description of the average community structure.  Inset: The structure of the eight blocks at the mutual information peak.}
\label{fig:MI}
\end{figure}

Figure~\ref{fig:MI} shows the results of this approach applied to our social network example.  The main plot shows the value of the reduced mutual information as a function of the number of blocks over the course of the calculation.  The plot has the expected form, with the value increasing to a maximum then falling off again.  The maximum value occurs for the case of eight blocks and the corresponding block structure is shown inset.

This choice of blocks does appear to be a good one.  Referring back to Fig.~\ref{fig:lesmis}, we show in the insets of panels B to~E a key that gives the mapping from blocks to communities for each of the structures depicted.  In each case it is possible to describe the entire community structure by saying to which community each block belongs, except for a small number of nodes, at most seven in any case (shown in bold), that do not fit the pattern.  Thus, while our community detection algorithm does indeed return a range of different divisions for this network, it is at the same time correct to say that those divisions all reflect essentially the same underlying structure in the network, since it is possible to express them as combinations of the same set of basic building blocks.

We note in passing an interesting feature of the blocks shown in Fig.~\ref{fig:MI}B: some of them are not connected, meaning they consist of two or more parts with no edges between parts.  This arises because our community detection algorithm is capable of finding disassortative structure in the network as well as assortative structure.  That is, it finds not only groups with a higher-than-expected number of edges, but also groups with a lower-than-expected number.  Some of the groups found in this case fall into the latter category and this is then reflected in the building blocks too.

\subsection{Further examples}
\begin{figure}
\begin{center}
\includegraphics[width=7cm]{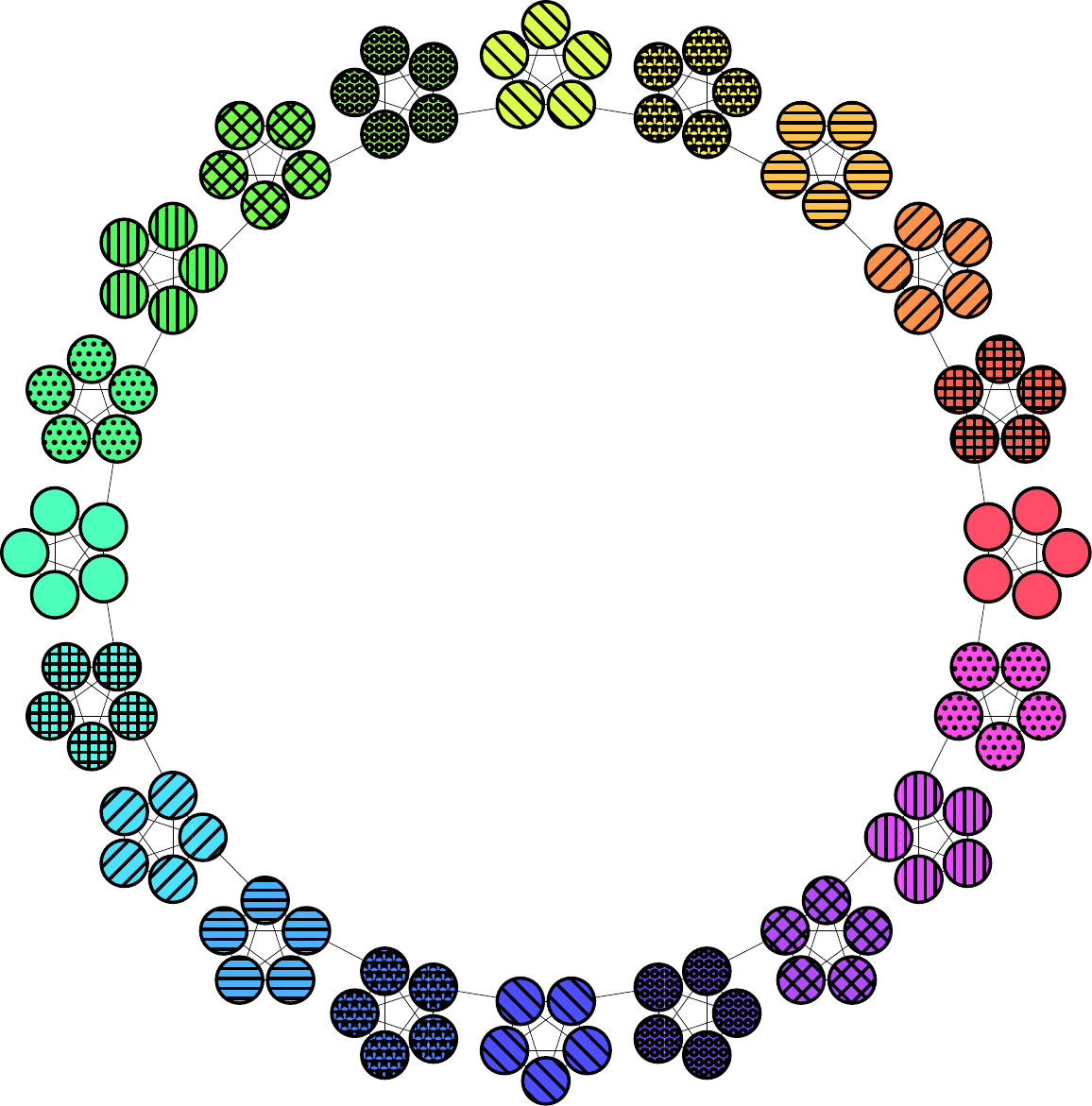}
\end{center}
\caption{Building blocks for the network of Fig.~\ref{fig:ring}, found by maximizing the reduced mutual information as described in Section~\ref{sec:MI}.  In this case the building blocks correspond exactly to the cliques within the network, as we would expect.}
\label{fig:cliques}
\end{figure}

Let us return to our first example, the ``ring of cliques'' network shown in Fig.~\ref{fig:ring}.  Our claim in Section~\ref{sec:cliques} was that the building blocks of this network were the cliques themselves, even though the communities found by the community detection algorithm are mostly larger than a single clique.  If this were true, and if the method of the previous section is indeed able to find the building blocks of a network, then when that method is applied to this network it should find the cliques.  And indeed it does.  Figure~\ref{fig:cliques} shows the optimal choice of building blocks for this network constructed using the greedy algorithm of the previous section and, as the figure shows, they correspond exactly to the 20 cliques in the network.  (We find equivalent results for networks with other numbers of cliques as well.)

\begin{figure}
\begin{center}
\includegraphics[width=\columnwidth]{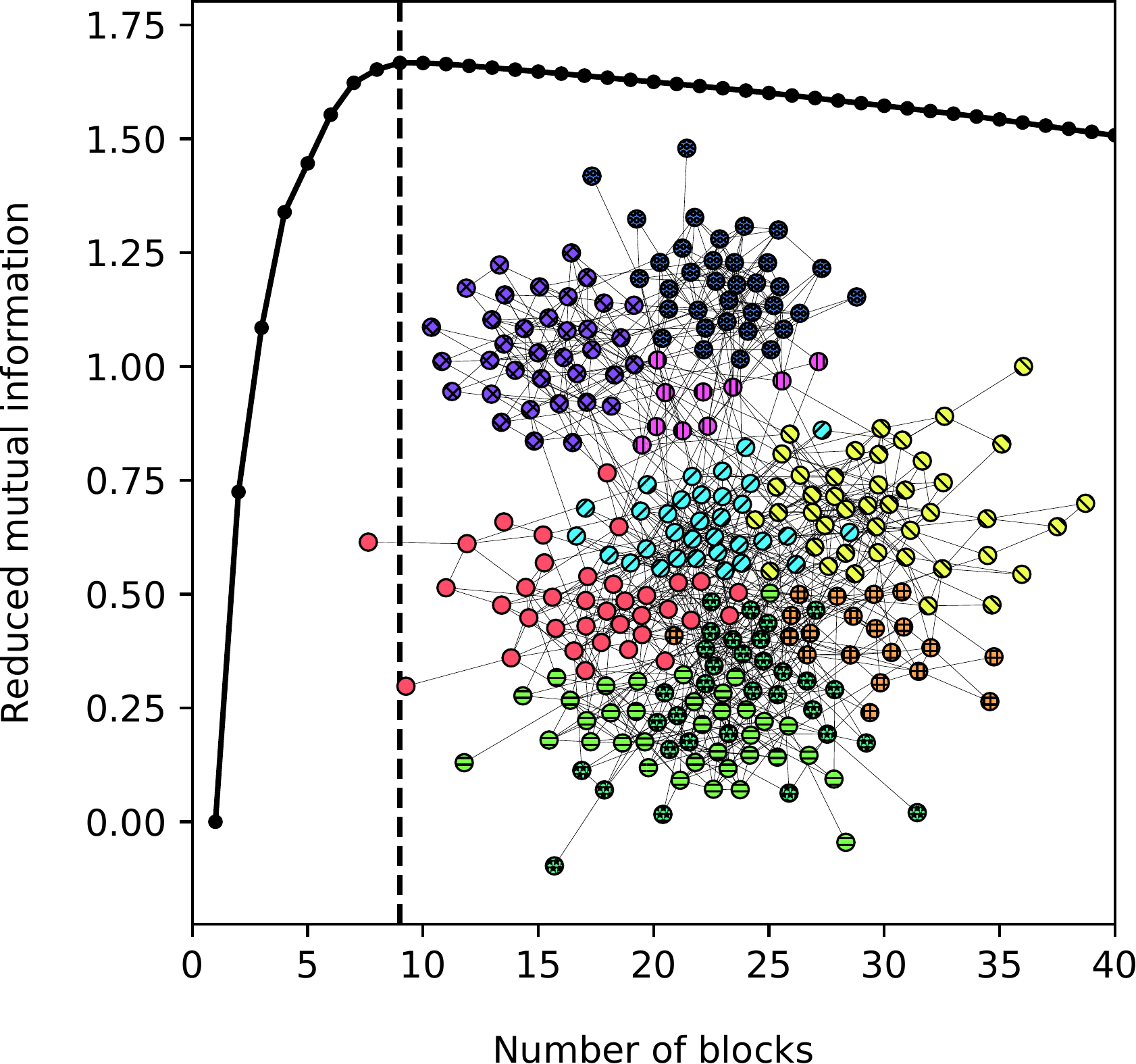}
\end{center}
\caption{Main figure: The reduced mutual information of Eq.~\eqref{eq:mi} for a social network of friendships among high-school students, averaged over $10\,000$ sampled community structures.  The mutual information has its maximum value when there are nine building blocks.  Inset: The structure of the blocks at the mutual information peak.}
\label{fig:addhealth}
\end{figure}

A more complex and realistic example is shown in Fig.~\ref{fig:addhealth}.  This network comes from the National Longitudinal Study of Adolescent to Adult Health (the ``Add~Health study''), a nationwide US social network study of friendship and dating behavior among middle- and high-school students (approximately ages 12 to 18 years).  The network pictured is a friendship network of self-identified friendships between students in one school out of the many that participated in the study.  (The school was picked primarily for its smaller size, which makes it easier to visualize the results.)  The main panel of Fig.~\ref{fig:addhealth} shows again the reduced mutual information as a function of number of blocks and for this network we see that the maximum value is reached for nine blocks.  Inset in the figure we show the corresponding set of blocks in the network, which in this case are all relatively compact sets of nodes.

\section{Conclusions}
In this paper we have examined community structure in complex networks using a fast Monte Carlo algorithm that samples high-likelihood structures.  We find, as a number of previous authors have also, that the typical network possesses good community divisions with a wide range of different structures.  We do not conclude, however, as some have done, that this indicates a failure either of the particular method of community detection or even of the entire community structure paradigm.  Instead, we observe that the competing structures are all related to one another in a relatively simple manner, namely they are all built from a small set of ``building blocks,'' groups of nodes that typically appear together in the same community.  The building blocks are not themselves communities in most cases, but complete community structures are formed by joining blocks together in various combinations.

We have argued that this implies that the amount of information needed to specify the community structure is small once the building blocks are known, and we use this fact to create an algorithm that can determine the blocks for any given network.  Starting from a large sample of plausible community structures generated by our Monte Carlo algorithm, we compute (a variant of) the mutual information between a proposed set of blocks and the community structure, averaged over all samples.  The optimal choice of blocks is the one that maximizes this average mutual information, which we find using an approximate greedy maximization algorithm.  We find that this algorithm can accurately recover the known blocks in a previously proposed class of test networks, and we also give example applications to real-world social networks.

The lesson behind these findings is that, while the existence of large sets of competitive and apparently disparate community structures in real and model networks appears at first to be a bad sign for community detection algorithms, the situation is actually a lot better than it seems.  The observed structures are, in essence, all variants of the same basic template, and the complete set of community divisions in fact provides significant information about the large-scale structure of the network.

\begin{acknowledgments}
The authors thank George Cantwell, Alec Kirkley, and Jean Gabriel Young for useful conversations.  This work was funded in part by the James S. McDonnell Foundation (MAR) and by the US National Science Foundation under grant DMS--1710848 (MEJN).
\end{acknowledgments}


\begin{thebibliography}{10}
\expandafter\ifx\csname url\endcsname\relax
  \def\url#1{\texttt{#1}}\fi
\expandafter\ifx\csname urlprefix\endcsname\relax\def\urlprefix{URL }\fi

\bibitem{GN02}
M.~Girvan and M.~E.~J. Newman, Community structure in social and biological
  networks. \textit{Proc. Natl. Acad. Sci. USA} \textbf{99}, 7821--7826 (2002).

\bibitem{Fortunato10}
S.~Fortunato, Community detection in graphs. \textit{Phys. Rep.} \textbf{486},
  75--174 (2010).

\bibitem{NG04}
M.~E.~J. Newman and M.~Girvan, Finding and evaluating community structure in
  networks. \textit{Phys. Rev. E} \textbf{69}, 026113 (2004).

\bibitem{Newman06b}
M.~E.~J. Newman, Modularity and community structure in networks. \textit{Proc.
  Natl. Acad. Sci. USA} \textbf{103}, 8577--8582 (2006).

\bibitem{BGLL08}
V.~D. Blondel, J.-L. Guillaume, R.~Lambiotte, and E.~Lefebvre, Fast unfolding
  of communities in large networks. \textit{J. Stat. Mech.} \textbf{2008},
  P10008 (2008).

\bibitem{HLL83}
P.~W. Holland, K.~B. Laskey, and S.~Leinhardt, Stochastic blockmodels: First
  steps. \textit{Social Networks} \textbf{5}, 109--137 (1983).

\bibitem{NS01}
K.~Nowicki and T.~A.~B. Snijders, Estimation and prediction for stochastic
  blockstructures. \textit{J. Amer. Stat. Assoc.} \textbf{96}, 1077--1087
  (2001).

\bibitem{KN11a}
B.~Karrer and M.~E.~J. Newman, Stochastic blockmodels and community structure
  in networks. \textit{Phys. Rev. E} \textbf{83}, 016107 (2011).

\bibitem{RB07b}
M.~Rosvall and C.~T. Bergstrom, An information-theoretic framework for
  resolving community structure in complex networks. \textit{Proc. Natl. Acad.
  Sci. USA} \textbf{104}, 7327--7331 (2007).

\bibitem{GSA04}
R.~Guimer\`a, M.~Sales-Pardo, and L.~A.~N. Amaral, Modularity from fluctuations
  in random graphs and complex networks. \textit{Phys. Rev. E} \textbf{70},
  025101 (2004).

\bibitem{MD06}
C.~P. Massen and J.~P.~K. Doye, Thermodynamics of community structure. Preprint
  cond-mat/0610077 (2006).

\bibitem{CMN08}
A.~Clauset, C.~Moore, and M.~E.~J. Newman, Hierarchical structure and the
  prediction of missing links in networks. \textit{Nature} \textbf{453},
  98--101 (2008).

\bibitem{RB06b}
J.~Reichardt and S.~Bornholdt, When are networks truly modular? \textit{Physica
  D} \textbf{244}, 20--26 (2006).

\bibitem{RL08}
J.~Reichardt and M.~Leone, ({U}n)detectable cluster structure in sparse
  networks. \textit{Phys. Rev. Lett.} \textbf{101}, 078701 (2008).

\bibitem{DKMZ11b}
A.~Decelle, F.~Krzakala, C.~Moore, and L.~Zdeborov\'a, Asymptotic analysis of
  the stochastic block model for modular networks and its algorithmic
  applications. \textit{Phys. Rev. E} \textbf{84}, 066106 (2011).

\bibitem{ZM14}
P.~Zhang and C.~Moore, Scalable detection of statistically significant
  communities and hierarchies, using message passing for modularity.
  \textit{Proc. Natl. Acad. Sci. USA} \textbf{111}, 18144--18149 (2014).

\bibitem{GDC10}
B.~H. Good, Y.-A. de~Montjoye, and A.~Clauset, Performance of modularity
  maximization in practical contexts. \textit{Phys. Rev. E} \textbf{81}, 046106
  (2010).

\bibitem{RB06a}
J.~Reichardt and S.~Bornholdt, Statistical mechanics of community detection.
  \textit{Phys. Rev. E} \textbf{74}, 016110 (2006).

\bibitem{RCRN17}
M.~A. Riolo, G.~T. Cantwell, G.~Reinert, and M.~E.~J. Newman, Efficient method
  for estimating the number of communities in a network. \textit{Phys. Rev. E}
  \textbf{96}, 032310 (2017).

\bibitem{Watts99a}
D.~J. Watts, \textit{Small Worlds}. Princeton University Press, Princeton
  (1999).

\bibitem{Knuth93}
D.~E. Knuth, \textit{The Stanford GraphBase: A Platform for Combinatorial
  Computing}. Addison-Wesley, Reading, MA (1993).

\bibitem{NCY19}
M.~E.~J. Newman, G.~T. Cantwell, and J.~G. Young, Improved mutual information
  measure for classification and community detection. Preprint arxiv:1907.12581
  (2019).

\end{thebibliography}
\end{document}